\documentclass[twoside]{article}
\usepackage{amsfonts,amssymb,amsmath}
\usepackage[dvips]{graphics,epsfig}

\begin{document}

\setcounter{equation}{0}

\begin{center}
{\large \bf THE RECEPTOR–TOXIN–ANTIBODY INTERACTION:

MATHEMATICAL MODEL AND NUMERICAL SIMULATION

}\vspace{4mm}

\textbf{\large P.\,Katauskis$^{1}$, P.\,Skakauskas$^{1}$, A.\,Skvortsov$^{2}$}\\[2mm]
\index{Katauskis P.}%
\index{Skakauskas V.}%
\index{Skvortsov A.}%

\emph{$^{1}$Vilnius University, Lithuania}\\
\emph{$^{2}$DSTO, VIC 3207, Melbourne, Australia}\\[1mm]

\medskip
E-mail: pranas.katauskis@mif.vu.lt, vladas.skakauskas@maf.vu.lt,

alex.skvortsov@dsto.defence.gov.au
\end{center}

{\def\thefootnote{}\footnotetext{\hspace{-15pt}\copyright\ \
 P.\,Katauskis, V.\,Skakauskas, A.\,Skvortsov 2012} }

\section{Introduction}

 \noindent An antibody, also known as an immunoglobulin,
 is a protein used by the immune system to identify,
neutralize, or kill foreign objects like bacteria, viruses, or pollen
which are termed as antigen. The production of antibodies is the
main function of the immune system. An antigen, when introduced into
the body, triggers the production of an antibody by immune system
which will then kill or neutralize the antigen that is recognized as
a foreign invader.

The bio-medical application of antibodies against an effect of
  toxins associated with various biological threats
 (epidemic outbreaks or bio-terrorism) is well-documented (see, e.g., [1--3]).

For a long time the main target of antibody design
 has been the antibody affinity. With progress in bio-engineering,
 many antibodies with different affinity parameters have been generated.
  However, according to Skvortsov and Gray [4] affinity is not a good
 predictor of protective or therapeutic potential of
  an antibody. In fact, the treatment effect of an antibody can be
    described by a parameter which includes the
    reaction rates of the receptor-toxin-antibody
      (RTA) kinetics and relative concentration of
      reacting species.
  As a result, any given value of this parameter determines a
       range of antibody kinetic properties and its
        relative concentration in order to achieve a
         desirable therapeutic effect.

The model considered by Skvortsov and Gray is a model of a well-mixed
solution of toxin, antibody, and cells and neglects diffusion fluxes
of interacting species. Diffusion fluxes are significant especially
when the process of RTA interaction is limited by diffusion.  Skakauskas
et al. [5] examined numerically a RTA interaction model taking into
account diffusion of all species in the case where a spherical cell
is embedded into an initially uniformly distributed toxin--antibody
 solution which occupies a large
volume (compartment) lying between the cell and external surface.
Initial values of species and their values on the external surface
were assumed to be the same for all times. In this case fluxes of
toxin, antibody and their complex across the external surface are
not zero. Some numerical results of the evaluation of an antibody
treatment efficiency parameter are given in this paper.

In the present paper by using the same model we study the influence
of RTA kinetic parameters and diffusivity of toxin, antibody,
and their complex on the behavior of the antibody protection
parameter and concentrations of species in more detail.

 The paper is organized as follows. In Section 2 we introduce the
 reaction--diffusion model for RTA interaction. Numerical
 results are presented in Section 3. Summarizing remarks given in Section 4 conclude the
 paper.

\section{The model}

 We study a case of a spherical
cell embedded into a toxin--antibody solution which occupies an
extracellular domain $\Omega$ lying between the cell and an external
surface and use notations of paper [5]:

  $\rho$  -- spherical radius,

 $S_c=\{\rho: \rho=\rho_c\}$ -- the surface of the spherical cell, $\rho_c$ is its radius,

 $S_e=\{\rho: \rho=\rho_e\}$ -- the surface of the external sphere (external surface of $\Omega$),
   $\rho_e$ is its radius,

$\Omega=\{\rho:\rho\in(\rho_c,\rho_e)\}$ -- the extracellular
domain,

 $r_0$ -- the concentration of receptors on the cell surface,

 $\theta(t,\rho)$ -- the fraction of the toxin-bound receptors,

 $r_0\theta$ -- the concentration of the toxin-bound
receptors (confined to $S_c$),

 $r_0(1-\theta)$ -- the concentration of the free receptors,

 $u_{T},$ $u_{A}$, and $u_{C}$ -- the concentrations of toxin, antibody, and
toxin--antibody complex, respectively,

 $ u_{T}^0,$ $u_{A}^0,$ $u_{C}^0$ -- the initial concentrations,

 $\kappa_T,$ $\kappa_A$, and $\kappa_C$ -- the diffusivity of the
toxin, antibody, and toxin--antibody complex, respectively,

 $k_1$, $k_{-1}$ -- the forward and reverse constants of the toxin--antibody
reaction rate,

 $k_2$ and $k_{-2}$ -- the forward and reverse constants of the toxin and receptor
binding rate,

 $k_3$ -- the  rate constant of the toxin internalization,

 $\partial_n$ -- the outward normal derivative on $S_e$ or $S_c$,

 $\partial_t=\partial/\partial t$,

 $\Delta=\rho^{-2} \dfrac{\partial}{\partial \rho}(\rho^2 \dfrac{\partial}{\partial \rho})$ --
 the Laplace operator,

 $\psi(t)$ -- the antibody protection factor (a relative reduction
of toxin inside a cell due to application of antibody).

Dynamics of the concentrations  $u_{T}, $ $u_{A}$, $u_{C}$, and
$\theta$  can be described by the following equations:

\begin{equation}
\begin{cases}\partial_t u_T=-k_1u_T
u_A+k_{-1}u_C+\kappa_T\Delta u_T,\quad \rho\in \Omega,\ t>0,
\\
 u_T=u_{T}^0,\quad \rho=\rho_e,\  t>0,\\
\partial_n u_T=\frac{r_0}{\kappa_T}(-k_2(1-\theta)u_T+k_{-2}\theta),\quad \rho=\rho_c,\
t>0,
\\ u_T\vert_{t=0}=u_{T}^0,\quad
\rho\in\Omega,
\end{cases}
\end{equation}

\begin{equation}
\begin{cases}\partial_t\theta=k_2(1-\theta)u_T-k_{-2}\theta-k_3\theta,\quad
\rho=\rho_c,\ t>0,
 \\ \theta\vert_{t=0}=0,\quad \rho=\rho_c,
\end{cases}
\end{equation}

\begin{equation}
\begin{cases}\partial_t u_A=-k_1u_T u_A+k_{-1}u_C+\kappa_A \Delta
u_A,\quad \rho\in\Omega,\ t>0,
\\ u_A=u_{A}^0, \quad \rho=\rho_e,\
t>0,
\\
 \partial_n u_A=0,\quad \rho=\rho_c,\ t>0,\\
u_A\vert_{t=0}=u_{A}^0,\quad \rho\in\Omega,
\end{cases}
\end{equation}

\begin{equation}
\begin{cases}\partial_t u_C=k_1u_T u_A - k_{-1}u_C+\kappa_C \Delta
u_C,\quad \rho\in\Omega,\ t>0,
\\ u_C=0,\quad \rho=\rho_e,\
t>0,
\\
\partial_n u_C=0,\quad \rho=\rho_c,\  t>0,
\\
u_C\vert_{t=0}=0,\quad \rho\in\Omega.
\end{cases}
\end{equation}

The initial and boundary conditions for the system above correspond
to a case where initially the toxin and antibody are distributed
uniformly in the extracellular domain $\Omega$. Values of all
species on the outer boundary of  $\Omega$ for all times and their
initial values are assumed to be the same.  In particular, zero
value of the toxin--antibody complex is used for initial time and
for all times on the outer boundary of  $\Omega$. We stress that in
this case the fluxes of all species are not zero on the outer
boundary $S_e$ of $\Omega$.

Eqs.~(1)--(4) can be presented in non-dimensional form by using
scales of $\tau_*$ (time), $l$ (length), and $u_*$ (concentration).
 By substituting variables
\medskip
 $x=l\bar x,$ $t=\tau_*\bar t,$ $r_0=lu_*\bar r_0$, $u_T=u_*\bar
u_T,$ $u_A=u_*\bar u_A,$ $u_C=u_*\bar u_C,$ $u_{T0}=u_*\bar
u_{T}^0,$ $u_{A0}=u_*\bar u_{A}^0$, $\bar k_1=\tau_* u_* k_1,$ $\bar
k_2=\tau_* u_* k_2,$ $\bar k_{-1}=\tau_* k_{-1},$ $\bar
k_{-2}=\tau_* k_{-2},$ $\bar k_3=\tau_* k_3$, $\bar\kappa_T=\tau_*
\kappa_T l^{-2},$ $\bar\kappa_A=\tau_*\kappa_A l^{-2},$
$\bar\kappa_C=\tau_*\kappa_C l^{-2}$
\medskip
into~(1)--(4) we can deduce the same system, but only in the
non-dimensional variables. Therefore, for simplicity in what
follows, we treat system (1)--(4) as non-dimensional.

The main antibody treatment efficiency parameter is the antibody
protection factor (a~relative reduction of toxin attached to a cell
due to application of antibody) which can be defined
 by the following expression~[4,5]:

\begin{equation}
\psi(t) =\frac{\int_{S_c}\theta\vert_{u_{A}^0 > 0}\,\mathrm{d}S }
{\int_{S_c}\theta\vert_{u_{A}^0=0}\,\mathrm{d}S}.
\end{equation}

By definition  $0 \leq \psi \leq 1$. The lower the value of $\psi$  the
more profound is therapeutic effect of antibody treatment.

\section{Numerical results}

\noindent We treated system~(1)--(4) numerically for the spherically
symmetric domain, $\rho\in(\rho_c,\rho_e)$, and $t>0$ with an
implicit finite-difference scheme. Our selection of the values of
parameters was motivated by the values available in the literature
[3,5--7] with the extended range to allow exploration and
illustration of the various transport and kinetics regimes that are
possible in the RTA system. We employ the following data that were
used in the most calculations in [5,8]:
 $u_*=6.02\cdot 10^{13}\ \mathrm{cm}^{-3},$ $\tau_*=1\ \mathrm{s},$
$r_0=1.6\cdot10^4/S_c$, where $1.6\cdot 10^4$ is the total number of
receptors of the cell, $l=10^{-2}\ \mathrm{cm},$ $S_c
=4\pi\rho_c^2=4\pi\cdot 10^{-6}\ \mathrm{cm}^2$, $\bar
r_0=2.115\cdot 10^{-3}$. The standard non-dimensional values of the
other parameters are the following:

\begin{equation}
\begin{cases} k_1=1.3\cdot 10^{-2},\quad k_{-1}=1.4\cdot 10^{-4},
\\ k_2=1.25\cdot 10^{-2}, \quad k_{-2}=5.2\cdot
10^{-4}, \quad k_{3}=3.3\cdot 10^{-5},
\\
\kappa_T=10^{-2}, \quad\kappa_A=10^{-2},\quad \kappa_C=10^{-2},
\\
\rho_c=10^{-1},\quad\rho_e=2,
\\u_A^0=1,\quad u_T^0=0.5.
\end{cases}
\end{equation}

These values correspond to the  ricin and 2B11 mono-clonal antibody
interaction. If values of $k_1$, $k_2$, $\kappa_A$, $\kappa_C$, and
$\kappa_T$ differ from those given in~(6), they are specified in the
legends of plots.

As we indicated in the Introduction, the main purpose of our study
was to estimate the effect of diffusive and kinetic parameters of
species on the behavior of concentrations of species and protective
properties of an antibody against a toxin. Results of numerical
solving of system~(1)--(4) are presented in Figs.~1--7.

\begin{figure}[!ht]
\begin{center}
\epsfxsize=0.5\textwidth\epsfbox{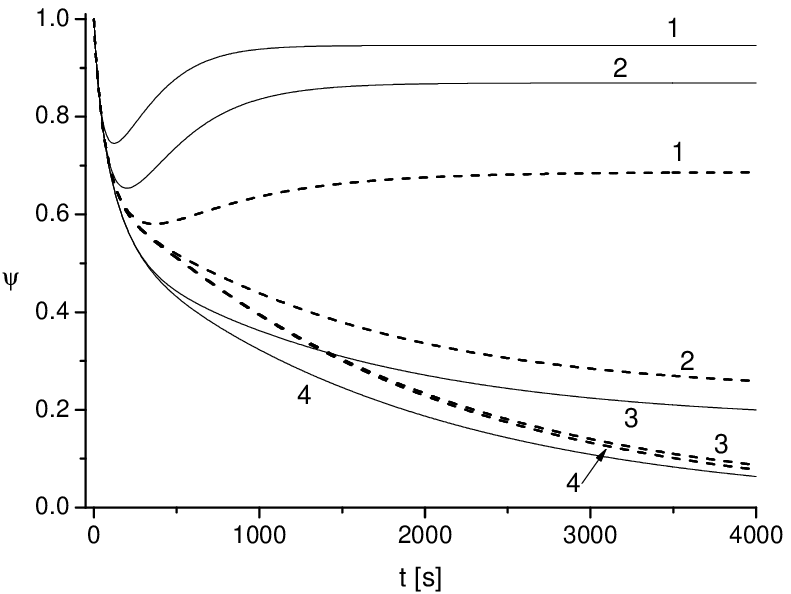}\\
{\small Fig. 1. Influence of the external radius $\rho_e=2$ (solid
line) and 5 (dashed line)
 and
  the toxin diffusivity $\kappa_T:$ $10^{-2}$ (1), $5\cdot 10^{-3}$ (2),
  $10^{-3}$ (3), $10^{-4}$ (4) on the cell
   protection characteristic, $\psi$, in the case of $u_T^0=0.6.$}
\end{center}
\vspace{-4mm}
\end{figure}

\begin{figure}[!ht]
\begin{center}
\epsfxsize=0.5\textwidth\epsfbox{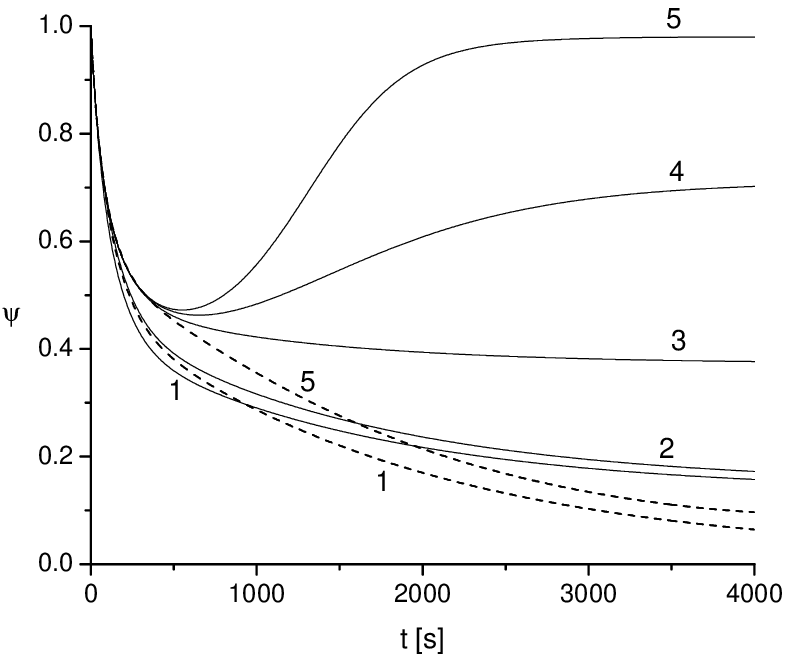}\\
{\small Fig. 2. Effect of the external radius $\rho_e=2$ (solid
line) and 5 (dashed line)
 and
  the antibody diffusivity $\kappa_A:$ $10^{-1}$ (1), $10^{-2}$ (2),
 $10^{-3}$ (3), $5\cdot 10^{-4}$ (4), $10^{-4}$ (5) on the cell
   protection factor, $\psi$,  in the
 case of $\kappa_T=10^{-3}.$}
\end{center}
\vspace{-4mm}
\end{figure}

\begin{figure}[!ht]
\begin{center}
\epsfxsize=0.5\textwidth\epsfbox{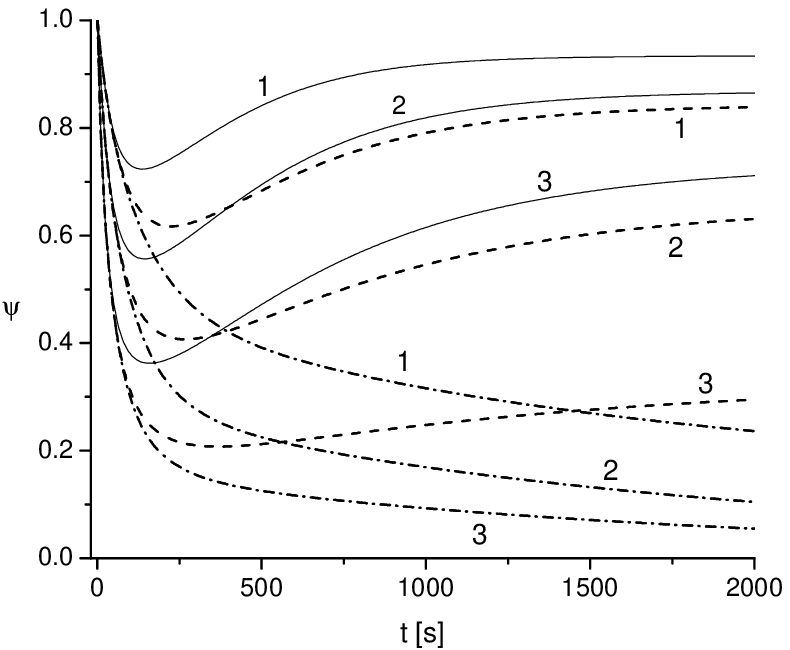}\\
{\small Fig. 3. Effect of the toxin diffusivity $\kappa_T:$ $
10^{-2}$ (solid line), $5\cdot 10^{-3}$ (dashed line), $10^{-3}$
(dash-dotted line) and parameter $k_1:$ $1.3\cdot 10^{-2}$ (1),
$2\times 1.3\cdot 10^{-2}$ (2), $4\times 1.3\cdot 10^{-2}$ (3)
 on the cell protection function $\psi.$
 }
\end{center}
\vspace{-4mm}
\end{figure}

\begin{figure}[!ht]
\begin{center}
\epsfxsize=0.5\textwidth\epsfbox{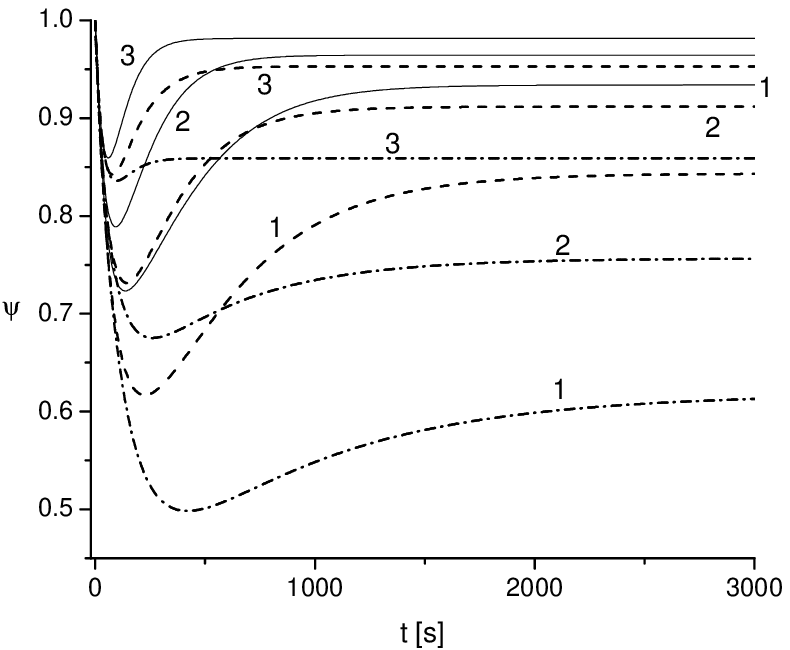}\\
{\small Fig. 4.  Effect of the toxin diffusivity $\kappa_T:$ $
10^{-2}$ (solid line), $5\cdot 10^{-3}$ (dashed line), $2.5\cdot
10^{-3}$ (dash-dotted line) and parameter $k_2:$ $1.25\cdot 10^{-2}$
(1), $2\times 1.25\cdot 10^{-2}$ (2), $4\times 1.25\cdot 10^{-2}$
(3)
 on the cell protection function $\psi.$}
\end{center}
\vspace{-4mm}
\end{figure}

\begin{figure}[!ht]
\begin{center}
\epsfxsize=0.5\textwidth\epsfbox{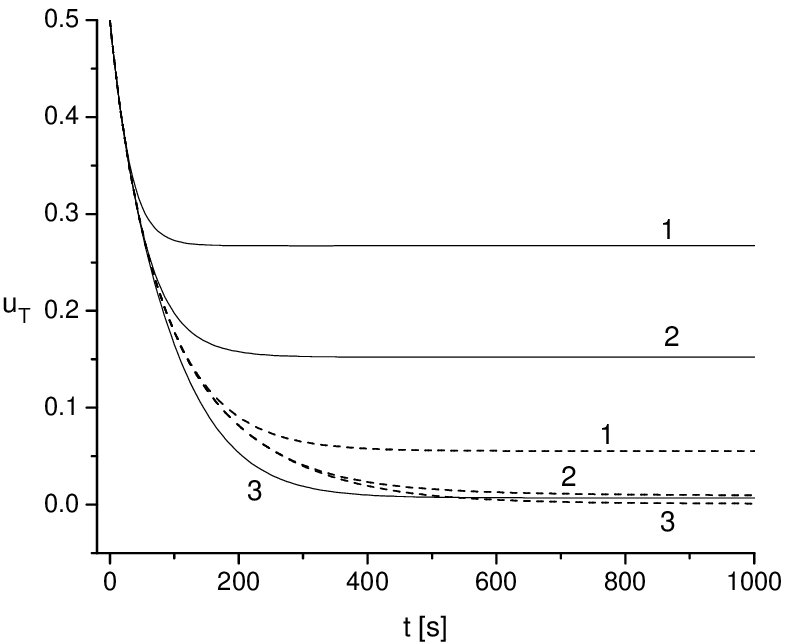}\\
{\small Fig. 5. Dynamics of toxin concentration $u_T$ for $\rho_e=2$
(solid line), $\rho_e=5$ (dashed line), and
   $\kappa_T:$ $10^{-2}$ (1), $5\cdot 10^{-3}$ (2), $10^{-3}$ (3). }
\end{center}
\vspace{-4mm}
\end{figure}

\begin{figure}[!ht]
\begin{center}
\epsfxsize=0.5\textwidth\epsfbox{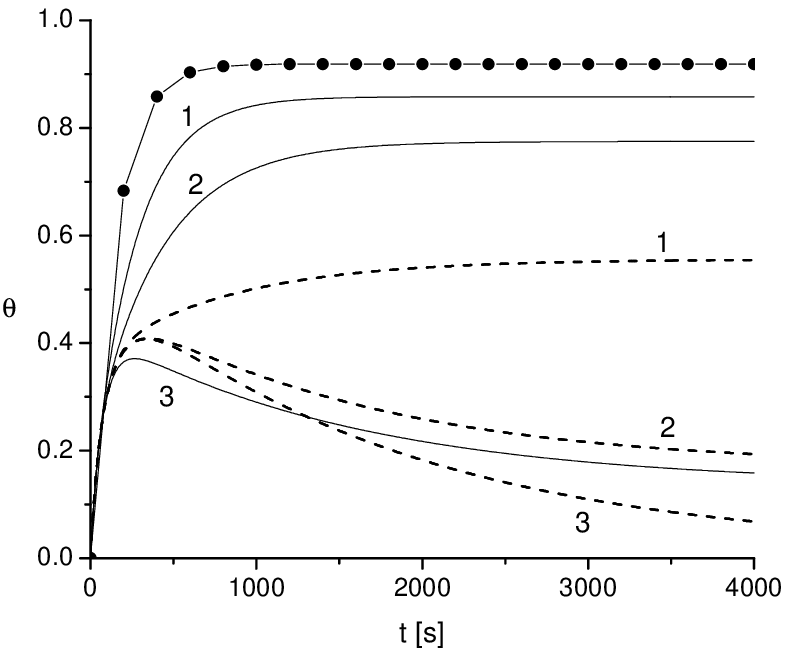}\\
{\small Fig. 6. Profiles of functions $\theta$ for $u_A^0=1$,
$\rho_e=2$ (solid line); $u_A^0=1$, $\rho_e=5$ (dashed line),
   and $\kappa_T:$ $10^{-2}$ (1), $5\cdot 10^{-3}$ (2), $10^{-3}$ (3).
   Line with bullets in the case of $u_A^0=0.$}
\end{center}
\vspace{-4mm}
\end{figure}

\begin{figure}[!ht]
\begin{center}
\epsfxsize=0.5\textwidth\epsfbox{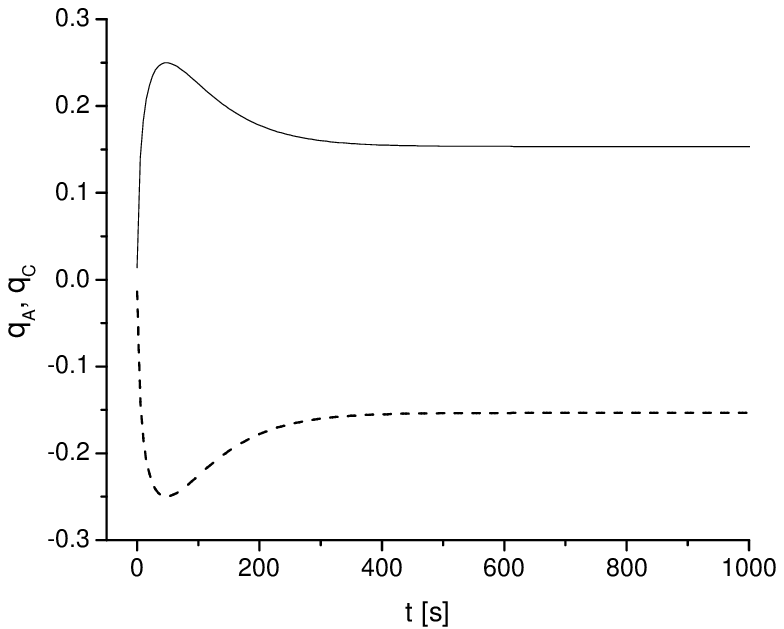}\\
{\small Fig. 7.  Dynamics of functions $q_A=\partial
u_A(t,\rho_e)/\partial \rho$ and $q_C=\partial
u_C(t,\rho_e)/\partial \rho$ at $\rho_e=2$ for $\kappa_T=10^{-3}$
 and $\kappa_A=\kappa_C=10^{-2}.$}
\end{center}
\vspace{-4mm}
\end{figure}

The plots of $\psi$ in Fig.~1 depict the dependence of the antibody
protection factor on the radius $\rho_e$ of the external surface
$S_e$ and toxin diffusivity $\kappa_T$. Parameter $\psi$ increases
with $\kappa_T$ growing, but its behavior for large values of
$\kappa_T$ is non-monotonic. For large values of $\kappa_T$,
parameter $\psi$ grows as $\rho_e$ decreases. But for small values
of $\kappa_T$ its behavior is different. For example, if
$\kappa_T=10^{-3}$, then values of $\psi$ for $\rho_e=5$ are larger
than those for $\rho_e=2$ if $t<1400$ s approximately. But if
$\kappa_T\le 10^{-4}$, then, for all $t$, values of $\psi$ for
$\rho_e=5$ are larger than those for $\rho_e=2$ (see curves 3
and~4).

 Fig.~2 illustrates the dependence of $\psi$ on the diffusivity  $\kappa_A$
 of the antibody. The curves in this figure depict the increase of
 $\psi$ as $\kappa_A$ decreases and non-monotonic time evolution of $\psi$
  for small values of $\kappa_A$. Moreover, in the case of small
 antibody diffusivity, $\kappa_A=10^{-4}$, values of $\psi$ for
 $\rho_e=2$ are larger than those for  $\rho_e=5.$ But in the case of
 large antibody diffusivity, $\kappa_A=10^{-1}$, values of $\psi$ for
 $\rho_e=2$ are smaller than those for $\rho_e=5.$ only if $t\le
 1000$ s. For $t>1000$ s they behave vica versa.

 Figs.~3 and 4 exhibit the dependence of $\psi$ on diffusivity
 $\kappa_T$,  forward constant $k_1$ of the toxin and antibody reaction rate, and forward constant
 $k_2$ of the toxin and receptor binding rate, respectively. Fig.~3
 demonstrates the decrease of $\psi$ as $k_1$  increases. But
 different values of $k_1$ do not change the monotonic behavior of all
 curves in time. From Fig.~4 we see the non-monotonic behavior of
 $\psi$ as $k_2$ increases. Moreover, $\psi$ increases with $k_2$
 increasing. The bottom of the hollow in Fig.~3 is located lower
 than that in Fig.~4. One can see in Fig.~4 that the effect of
 toxin diffusivity variation on protection factor is sensitive
 to changes of parameter $k_2$. Let
us compare the minimal values of protection factor. In the case of
$k_2=1.25\cdot 10^{-2}$, the minimum of $\psi$ is about 0.72 at
$\kappa_T=10^{-2}$, 0.62 at $\kappa_T=5\cdot 10^{-3}$ and 0.5 at
$\kappa_T=2.5\cdot 10^{-3}$, while the corresponding values of
$\psi$ are about 0.86, 0.84 and 0.836 in the case of $k_2=5\cdot
10^{-2}$ (curves 1 and~3).

 Numerical experiments show that diffusivity $\kappa_C$
 practically does not influence the time evaluation of $\psi$.

 The plots of $u_T$ in Fig.~5 depict the dependence of the toxin
 concentration at the cell surface on the diffusivity $\kappa_T$ and radius $\rho_e$
 of the external surface $S_e$. For any value of $\rho_e$, $u_T$ decreases with $\kappa_T$
 decreasing. For large values of $\kappa_T$, function $u_T(t,\rho_c)$
grows as $\rho_e$ decreases. But, for small values of $\kappa_T$,
its behavior is different. For example, for $\kappa_T=10^{-3}$,
values of $u_T(t,\rho_c)$ for $\rho_e=5$ are larger than those for
$\rho_e=2$ only if $t<500$ s (see curves 3).  Our calculations show
that influence of $\kappa_C$ on the behavior of $u_T(t,\rho_c)$ is
insignificant. We observed the non-monotonic behavior of
$u_T(t,\rho_c)$ for small  $\kappa_C$, but difference between its
steady-state value and value at the bottom of the hollow is very
small (of order $10^{-3}$).

Calculations show that $u_C(t,\rho_c)$ grows  with  $\kappa_C$
decreasing. The behavior  $u_C(t,\rho_c)$  is monotonic for
$\kappa_C\le 5\cdot 10^{-2}$. Its values are smaller than initial
ones of toxin for $\kappa_C\in [5\cdot 10^{-3},5\cdot 10^{-2}]$. But
$u_C(t,\rho_c)$ can reach a relatively large steady-state value for
small $\kappa_C$ while steady-state values of $u_T$ and $u_A$ are
smaller than their initial values.
 For example, the steady-state value of $u_C$ on $S_c$ is equal to $2.2$
for $\kappa_C=10^{-3},\,\kappa_A=\kappa_T=10^{-2}$. Derivatives of $u_T$ and $u_A$ with respect to $\rho$
on $S_e$ are of order $0.3$ while derivative of $u_C$ on $S_e$ is of order $-3$. This means that $u_C$ increases faster towards the cell than $u_T$ and $u_A$ decays in the same direction.

Curves in Fig.~6 depict the dependence of $\theta$ on $\kappa_T$ and
$\rho_e$ for $u_A^0=1$ and $u_A^0=0$. In the case where the antibody
is absent values of $\theta$ practically do not depend on
diffusivity $\kappa_T$ (see the bullets marked curve).
$\theta$~decreases with $\kappa_T$ decreasing. For any $\rho_e$,
function $\theta$ grows as $\kappa_T$ increases.
 If $\kappa_T\in 5\cdot [10^{-3},10^{-2}]$, then values of $\theta$ for $\rho_e=2$
 are larger than those for $\rho_e=5$. But for small values of $\kappa_T$ its behavior is
different. For example, if $\kappa_T=10^{-3}$, values of $\theta$ for
$\rho_e=5$ are larger than those for $\rho_e=2$ only for about
$t<1300$ s. This behavior is similar to those of $u_T$ and $\psi$.

Two curves in Fig.~7 illustrate the non-monotonic behavior of
derivatives $\partial u_A(t,\rho_e)/\partial \rho$  and $\partial
u_C(t,\rho_e)/\partial \rho$ for small toxin diffusivity
($\kappa_T=10^{-3}$). For $\kappa_T=10^{-2}$ their behave is
monotonic.

\section{Concluding remarks}

To conclude the paper we summarize results of study. The
receptor--toxin--antibody interaction is studied numerically by
using a model proposed in [5]. The model includes "bulk" reaction of
 toxin and antibody, surface binding of toxin and
 cell receptors, and diffusion of all species.
  The main results of the numerical study are the following:

1. The evolution of concentrations of some species
(toxin and toxin-bound receptors) and of the antibody protection factor
for some cases (large toxin diffusivity, small antibody diffusivity,
and large forward constant of the toxin--receptor binding rate) is non-monotonic

2. The influence of small or large values of $\kappa_T$, $\kappa_A$,
and $k_2$ on the behavior of $u_T(t,\rho_c),$ $\theta(t)$ and
$\psi(t)$ is profoundly different in the cases of small or large
$\rho_e$.

 3. The effect of $\kappa_C$ on the evolution of $u_T,$ $\theta$,
and $\psi$ was found to be insignificant.





\bigskip

\noindent {\small \textbf{R~e~f~e~r~e~n~c~e~s}}\vspace{-2mm}

\begin{enumerate}
\itemsep=-0.7mm

{\small%

\item \emph{Oral H.B., Ozakin C., Akdis C.A.} Back to the future: antibody-based
strategies for the treatment of infectious diseases // Mol.
Biotechnol. 2002. T.~21. P.~225--239.

\item \emph{Lobo E.D., Hansen R.J., Balthasar J.P.} Antibody
pharmacokinetics and pharmacodynamics // J. Pharm. Sci. 2004. T.~93.
P.~2645--2668.

\item \emph{Prigent J., Panigai L., Lamourette P., Sauvaire D., Devilliers K. et
al.} Neutralising antibodies against ricin toxin // PloS ONE. 2011.
T.~6. P.~e20166.

\item \emph{Skvortsov A., Gray P.} Modeling and simulation of
receptor--toxin--antibody interaction // Proc. 18th World IMACS/
MODSIM Congress. Cairns, Australia, 2009. P.~185--191.

\item \emph{Skakauskas V., Katauskis P., Skvortsov A.} A reaction--diffusion model of
the receptor--toxin--antibody interaction // Theor. Biol. Med.
Model. 2001. T.~8:32. P.~1--15.

\item \emph{Sandvig K., Olsnes S., Pihl A.} Kinetics of binding of the toxic
lectins abrin and ricin to surface receptors of human cells // J.
Biol. Chem. 1976. T.~251. P.~3077--3984.

\item  Lectures Notes in Immunology: Antigen--antibody interactions,
University of Pavia.

 $\mathrm{http://nfs.unipv.it/nfs/minf/dispense/immunology/lectures/files/antigens\_antibodies.html}$
2011.

\item \emph{Truskey G.A., Yuan F., Katz D.F.}  Transport Phenomena in
Biological Systems, second ed. Prentice Hall, 2009. 888 p.


}
\end{enumerate}

\end{document}